\documentclass[12pt]{JHEP3}

\usepackage{amssymb, amsmath, amsopn, amsthm}
\usepackage{epsfig}

\newcommand{\Exp}[1]{\left\langle~#1~\right\rangle}

\newcommand{\bea}{\begin{eqnarray}}
\newcommand{\eea}{\end{eqnarray}}

\newcommand{\be}{\begin{equation}}
\newcommand{\ee}{\end{equation}}

\newcommand{\p}{\partial}

\title{Can a stationary Bianchi black brane have momentum along the direction 
with no translational symmetry? 
}

\author{Norihiro Iizuka$^{1}$, Akihiro Ishibashi$^{2}$ and Kengo Maeda$^{3}$~\\
 
$^1$Interdisciplinary Fundamental Physics Team, 
Interdisciplinary \\ 
Theoretical Science Research Group,  RIKEN, Wako 351-0198, JAPAN 
\thanks{Address after April 1, 2014: {\itshape\footnotesize{Department of Physics, 
Osaka University, 
560-0043, JAPAN}}}  
\\
\email{norihiro.iizuka@riken.jp} \\

$^2$Department of Physics, Kinki University, Higashi-Osaka 577-8502, JAPAN \\
\email{akihiro@phys.kindai.ac.jp} \\

$^3${\it Faculty of Engineering,
Shibaura Institute of Technology, \\ Saitama 330-8570, JAPAN} \\
\email{maeda302@sic.shibaura-it.ac.jp}

}

\abstract{
Bianchi black branes (black brane solutions with homogeneous but anisotropic horizons classified by the Bianchi type) provide a simple holographic setting with lattice structures taken into account. In the case of holographic superconductor, we have a persistent current with lattices. Accordingly, we expect that in the dual gravity side, a black brane should carry some momentum along a direction of lattice structure, where translational invariance is broken. Motivated by this expectation, we consider whether---and if possible, in what circumstances---a Bianchi black brane can have momentum along a direction of no-translational invariance. First, we show that this {\it cannot} be the case for a certain class of stationary Bianchi black brane solutions in the Einstein-Maxwell-dilation theory. Then we also show that this {\it can} be the case for some Bianchi VII$_0$ black branes by numerically constructing such a solution in the Einstein-Maxwell theory with an additional vector field having a source term. The horizon of this solution admits a translational invariance on the horizon and conveys momentum (and is ``rotating'' when compactified). However this translational invariance is broken just outside the horizon. This indicates the existence of a black brane solution which is regular but non-analytic at the horizon, thereby evading the black hole rigidity theorem. 
}

\begin{document}
\tableofcontents

\section{Introduction}\label{sec:intro}

Recent developments of revealing a rich class of new IR structure of black 
holes/branes show that various different classes of field theory dynamics 
could be realized in black hole/brane solutions in gravity side. 
One of these examples is holographic superconductor/superfluidity 
\cite{Gubser:2008px, Hartnoll:2008vx, Hartnoll:2008kx, Herzog:2009xv, Horowitz:2010gk, Horowitz:2009ij}. 
Symmetry breaking in the IR in gauge theory is reflected as 
various new hairy black brane solutions in gravity. 
Another interesting black brane solutions  
are Lifshitz geometry \cite{lifsol1,lifsol2,lifsol3} and so-called geometry with hyperscaling violation \cite{hv1,Iizuka:2011hg,hv31,hv3,hv32,hv4,hv5,hv6,Edalati:2012tc,Bueno:2012vx}. 
There, geometry shows more anisotropic solution in the sense that scaling dimension is more generic than Lorentz-invariant case,  
but respects spatial rotation and translational invariance in the IR. Therefore these can be  
bulk dual to homogeneous and isotropic but not necessary Lorentz invariant field theory.   
Furthermore, these geometries 
can admit vanishing entropy density at the 
zero temperature limit, which is natural from thermodynamical point of view.   
As an attempt to explore various IR geometries, the 5-dimensional extremal black brane geometries which admit homogeneous but anisotropic in IR are classified using 
the so-called 3-dimensional ``Bianchi''-classification~\cite{IizukaBianchiI}. 
More generic exotic brane classification, using 4-dimensional extension of 
3-dimensional Bianchi-classification, was done in 3-dimensional 
Bianchi-classification \cite{IizukaBianchiII}. 
These correspond to the vacuum classification in the boundary gauge theories.

These Bianchi black branes are interesting since they do not in general 
admit translational invariance, but rather admit non-commutative three 
Killing vectors $\xi_i$ that obey nontrivial Lie algebra. 
Since $\partial_{x^i}$, a spatial shift translation along $x^i$, 
does not necessarily coincide a Killing vector except Bianchi I, 
the geometry does not in general have a translational invariance 
along the direction of $x^i$. 
An especially interesting class is of type VII$_0$, in which the 
three-dimensional homogeneous space forms a periodic ``helical structure''.  
As pointed out by \cite{Donos:2012js}, the directions with no translational 
invariance can be regarded as ``latticed directions''. 
Lattice effects which break the translational invariance are especially 
interesting and important in realistic condensed matter systems 
because momentum dissipation occurs due to the lattice. 
Quite recently, we have shown that there exists a new stationary 
hairy Bianchi VII$_0$ black brane solution dual to a superconducting state 
in the strongly coupled field theory side and the angular momentum 
exists along the ``latticed direction''~\cite{Iizuka2013Dec}~(See also 
the recent analysis of lattice effects in holographic superconductors 
\cite{Horowitz:2013, Iizuka:2012dk}). This implies that persistent current exists along the latticed direction and its resistance becomes zero even though 
there is no translational invariance along the direction of the current.

From general relativistic point of view, it is also interesting to consider 
in what circumstances a black brane can have momentum (if possible) 
along a direction with no translational symmetry. 
In the asymptotically flat case, one can speculate what would happen 
when an {\it asymmetric} black hole rotates: 
It would emit gravitational waves that carry angular momentum 
away, and eventually the rotation of the black hole would be damped out and 
the geometry would approach a static solution. 
In other words, as long as it is rotating, an asymmetric black hole 
will never be exactly stationary. 
This view is closely related to a consequence of the black hole rigidity theorem 
\cite{HE73, HIW07, MI08} that a stationary rotating black hole must be 
axisymmetric. However, in the asymptotically AdS black hole case, which is 
more relevant to the holographic context, the emitted gravitational radiation 
would be reflected back by the AdS infinity and its backreaction could 
possibly make the geometry either a state of forever 
dynamical (see, e.g.,~\cite{BR11,MR13, BLL13, DHMS12}) or an equilibrium 
state with no axisymmetry due to the presence of gravitational waves outside 
the horizon, besides the possibility that the geometry would settle down to 
another stationary, axisymmetric solution.   
In particular, for the equilibrium, non-axisymmetric case, the event horizon 
itself does not rotate with respect to the generator of the stationary 
symmetry according to the rigidity theorem, but instead some radiation outside 
the horizon would carry the angular momentum, presumably making the bulk 
geometry non-static. 
If some matter fields are included, they could also be a carrier of 
(part of) the angular momentum. (This is in fact the case in our previous 
paper~\cite{Iizuka2013Dec}. See also an earlier result along this line 
\cite{DiasHorowitzSantos2011}). 
Furthermore, in the context of gauge/gravity duality, we are typically 
concerned with black branes with {\it non-compact} horizon, 
to which the standard form of the rigidity theorem 
no longer applies, as the {\it compactness} of horizon cross-sections 
is one of the essential requirements for the proof of the theorem.  
(See also references \cite{Figueras:2012rb, Fischetti:2012vt} 
for constructions of black branes that evade 
the rigidity theorem by considering non-compact horizons.) 
It is therefore worth considering what would happen when a black brane, 
whose horizon cross-sections are not necessarily compact, gains momentum 
along a direction of no translational symmetry. For this line of study, 
the Bianchi black brane models described above provide a good starting point 
since, as we will see later, the equations of motion for the Bianchi models 
reduce to a set of ordinary differential equations.

The purpose of this paper is to reveal some non-trivial relation between 
the symmetry and momentum of a certain class of Bianchi black brane models 
that are expected to be dual to strongly coupled quantum systems, such as 
superconductors with lattices. 
We will supply a new example of Bianchi black brane solutions in 
the Einstein-Maxwell theory with an additional vector field. 
In our previous paper~\cite{Iizuka2013Dec}, we have constructed a stationary 
hairy Bianchi VII$_0$ black brane solution, in which the horizon is not 
rotating but matter fields carry momentum. 
The point of our construction in~\cite{Iizuka2013Dec} is 
that a $U(1)$ gauge field acquires a mass, which is a desired phenomenon 
in accord with a superconducting state in the dual theory side. 
Then one might expect that a Bianchi black brane cannot carry momentum 
unless a gauge field of the model becomes massive. 
We will show that this is indeed the case in the same type of Bianchi black 
branes considered in~\cite{Iizuka2013Dec} 
in Einstein-Maxwell and Einstein-Maxwell-dilaton theories.

We shall also explore another possibility that a black brane can have 
momentum along a direction of no translational invariance even 
in the case that a gauge field involved does not gain a mass. 
We will numerically construct such an example for black branes of 
Bianchi VII$_0$ type, by considering a model with a source term 
in the bulk Lagrangian. 
In this solution, the horizon can be compactified and is in fact 
``rotating'' (when compactified) along the direction of translational 
invariance on the horizon and hence conveys momentum. 
However, this translational invariance is actually violated just outside 
the horizon. 
This would be impossible if the metric and fields are {\it analytic} 
in a neighbourhood of the horizon, since in that case, the geometry---in 
particular, its symmetry property---near the horizon can be uniquely 
extended into the entire region by analytic continuation (under an additional 
condition of simple connectedness).  
Therefore this solution indicates, within the accuracy of our numerical method, 
the existence of a black brane which is regular but is not analytic  
at the horizon. The analyticity is a crucial requirement for the proof of 
rigidity theorem, besides the compactness of the horizon cross-sections. 
Therefore, although a mathematically more rigorous proof for this to be truly 
so is desired, our solution provides a qualitatively new, intriguing example 
of regular black branes which possess a stationary Killing vector field 
tangent to the horizon but nevertheless evade the rigidity theorem 
by the violation of analyticity.

The organization of this paper is as follows.  
In the next section, we show a theorem concerning the decomposition of 
Killing symmetry on the event horizon in a class of stationary Bianchi black 
branes. In section {\ref{section3}}, we show that 
Einstein-Maxwell theory or Einstein-Maxwell-dilaton theory cannot have 
a rotating black brane solution with momentum/rotation along the direction 
of no translational invariance for the same class of stationary Bianchi black 
branes considered in section~{\ref{section2}}. 
This corresponds to the lack of persistent current in normal phase 
(non-superconductor) along the direction of no translational invariance. 
In section {\ref{section4}}, we illustrate a model with a source term in the bulk Lagrangian which 
generates an explicit stationary solution where the horizon is rotating along 
the direction of no translational invariance except the horizon. 
Section {\ref{section6}} is devoted to summary and discussion.

\section{Symmetries on the horizon}
\label{section2}
In this section and the next, we show that the event horizon in a certain 
class of stationary 5-dimensional Bianchi black branes cannot rotate along 
the direction of no translational invariance. 
As discussed above, this is reminiscent of the black hole rigidity 
theorem. However it should be noted that the rigidity theorem requires, 
in an essential way, the compactness of the horizon cross-sections. 
In contrast, we will here deal with stationary black brane whose horizon 
cross-sections are not necessarily compact. 
In this section we discuss symmetry aspects focusing on the horizon, 
without any restriction to the type of theories, other than the requirement of 
null convergence condition. In the next section we turn our attention 
to the exterior of the horizon and consider momentum carried by matter fields 
within Einstein-Maxwell(-dilaton) theory.

In what follows, by 5-dimensional stationary Bianchi black brane we mean a 
stationary geometry containing a black hole, whose exterior region can 
be foliated by 4-dimensional timelike homogeneous hypersurfaces ${\cal N}(r)$, 
labeled by a parameter $r$ and spanned by mutually independent 
four Killing vector fields, $k$ and $\xi^I$ ($I=1,2,3$). 
Here $k$ is an asymptotically timelike Killing vector field 
with complete orbits, 
and $\xi^I$ are assumed to be everywhere spacelike, commute with $k$, 
and satisfy the Lie algebra 
\bea 
[\xi_I,\, \xi_J]={C^K}_{IJ}\xi_K
\eea
in the Bianchi classification~(see Appendix A in detail). 
Then, with the foliation parameter $r$ suitably chosen as the radial 
coordinate and the Killing parameter $t$ of $k=\p_t$ as the time 
coordinate, the 5-dimensional metric is written as 
\bea 
\label{metric_tr}
ds_5^2=-h\, dt^2+\frac{dr^2}{f}+2\tilde{N}_I dt\, \omega^I +\tilde{g}_{IJ}\omega^I \omega^J, \quad I,\,J=1,2,3\,, 
\label{def:metric}
\eea
where $h$, $f$, $\tilde{N}_I$, and $\tilde{g}_{IJ}$ are functions of $r$ only, 
and each $\omega^I$ denotes the invariant one-form associated with $\xi_I$.  
When the shift vector $\tilde{N}_I$ is non-vanishing, 
the spacetime can possess momentum/rotate with respect to the asymptotic 
observers along the orbits of $k$. 
When the horizon is rotating, $h$ is negative since 
$k$ becomes spacelike on the horizon, i.e.,~$g_5(k,\,k)>0$.  
Let us denote the location of the horizon $H$ by $r=r_h$~($f(r_h)=0$). 
Then, the horizon $H$ consists of a homogeneous null hypersurface spanned 
by $\xi^I$ and $k$ in the limit, $r\to r_h$, of $\cal N$. 
Now choose a horizon cross-section $\Sigma$ of $H$ in a way that 
on $H$, $k$ is everywhere transverse to $\Sigma$. 
Then Lie-dragging $\Sigma$ by $k$ defines a foliation $\Sigma(u)$ of $H$, 
where the parameter $u$ is given so that ${\cal L}_k u=1$.  

\medskip \noindent 
{\bf Theorem.~1} 
Consider a 5-dimensional stationary black brane with the stationary Killing 
vector $k$, and choose a horizon cross-section $\Sigma$ in the manner 
described above. Then $k$ is uniquely decomposed into a null vector field 
$n$ normal to $H$ and a spacelike vector $\eta$ lies in $\Sigma$ as 
\bea
k=n-\eta \,.   
\eea
Now suppose $\Sigma$ admits the Bianchi symmetry of the type II, VI$_0$, 
or VII$_0$. Then, under the null energy condition, either $\eta=0$, or 
$\eta$ must be a Killing vector of $\Sigma$, irrespective of $\Sigma$ 
being compact or non-compact.  
\\ 

\medskip \noindent 
{\it Proof}.   
Since $k$ generates an isometry, it must be tangent to $H$. 
When $k$ is normal to $H$, $\eta$ is identically zero. 
So, hereafter we focus on the case that $k$ is spacelike on $H$ 
and hence $N^2:=g_5(k,\, k)>0$ on $H$. 
Since the Killing vectors $k$ and $\xi^I$ all tangent to $H$, 
the metric of $H$ may be written by a local coordinate $u$ with 
${\cal L}_k u=1$ as 
\bea
ds_4^2=N^2du^2+2NN_I \omega^I du+g_{IJ}\,\omega^I \omega^J, 
\eea
where $N$, $N_I$, and $g_{IJ}$ are constants satisfying $\det g_{IJ}>0$ 
(in addition to $N^2>0$ mentioned above) and where, furthermore, 
the determinant of the gram matrix of $k$ and $\xi_I$ is required to vanish. 
If one expands $\eta$ by the invariant vectors $X_I$ 
as $\eta=\alpha^I\,X_I$, then the coefficients $\alpha^I~(=g^{IJ}\alpha_J)$ 
is uniquely determined by 
\bea 
\alpha_I=-N\,N_I, \qquad N_I\,N^I=1 
\eea
by the conditions $g_4(n, \,n)=0$ and $\p_{\alpha_I}g_4(n,\,n)=0$, 
as $H$ is an achronal surface~\footnote{An achronal surface is 
a surface which contains no two points with timelike separation. 
The only non-spacelike curve that can lie in $H$ is a null geodesic on $H$.}.  
It is straightforward to check that the gramian in fact vanishes.

Now let us consider a symmetric tensor $B_{ij}$ defined by 
$B_{ij}:=\widehat{\nabla_i l_j}~(i,\,j=1,2,3)$ with $l$ being 
the tangent vector for the null geodesic generator of $H$ 
with an affine parameter $\lambda$, 
where the hat means the projection onto the spatial cross-section 
$\Sigma$~\footnote{Since $l$ is 
a hypersurface orthogonal, the anti-symmetric part of $B_{ij}$ 
automatically vanishes.}, and here and hereafter the indices 
$i,\,j=1,2,3$ denote local coordinates $x^i$ on $\Sigma$. 
Then, the expansion $\theta$ and the shear tensor $\sigma_{ij}$ are defined by 
\bea 
\theta={B_i}^i, \quad 
\sigma_{ij}=B_{ij}-\frac{1}{3}\theta g_{ij}=\frac{1}{2}{\cal L}_l g_{ij}-\frac{1}{3}\theta g_{ij}, 
\eea
where $g_{ij}$ is the induced metric of $\Sigma$. 
By rewriting $l$ as $l=f\, n$ with some appropriate function $f$, 
the expansion $\theta$ is calculated as 
\bea 
 & \theta=g^{ij}B_{ij} 
         = \displaystyle\frac{1}{2} f g^{ij}{n}^\mu \p_\mu g_{ij} 
           +f\alpha^I\p_i {X_I}^i \,, 
\eea
where we used $g_4(n,\,\p_i)=0$. 
Note that if the horizon cross-section is compact, hence the area 
of $\Sigma$ is well-defined, then one can immediately show 
that $\theta=0$ under the null convergence condition by exploiting 
type of arguments used for the area-theorem~\cite{HE73}. 
However we do not proceed in that way, as we are interested also 
in the case where the horizon cross-sections are non-compact.

Now here we use some particular property of the Bianchi 
type II, VI$_0$, VII$_0$: 
For those, we find that the divergence, $\p_i {X_I}^i $, of the invariant 
vector $X_I$ vanishes for an arbitrary index $I$. 
Using this fact, we then find that 
\bea
\theta
 = \displaystyle\frac{1}{2}fg^{ij}{n}^\mu \p_\mu g_{ij} 
 = \displaystyle\frac{1}{2A}\frac{dA}{d\lambda} \,,    
\eea
where $A=\det (g_{ij})$. 
In terms of the local coordinates $x^i$, $g_{ij}$ on $\Sigma$ is given by 
\bea 
& ds_3^2=\widehat{g_{IJ}}\,\omega^I(x) \omega^J(x)=g_{ij}(x)dx^i dx^j, \quad I,\,J=1,2,3, \nonumber \\
& g_{ij}(x)=\widehat{g_{IJ}}\,{\Lambda^I}_i(x) {\Lambda^J}_j(x), \qquad \omega^I(x):={\Lambda^I}_i(x) dx^i. 
\eea
Along each null geodesic congruence of $l$, 
the Raychaudhuri equation holds: 
\bea 
\frac{d\theta}{d\lambda}=-\sigma_{ij}\sigma^{ij}-R_{\mu\nu}l^\mu l^\nu-\frac{1}{3}\theta^2. 
\eea
Here we use again a particular property of Bianchi II, VI$_0$, VII$_0$: 
For those types, one can find that besides the fact that $\widehat{g_{IJ}}$ is 
a constant matrix on $H$, the determinant $\det \Lambda^I{}_i(x)$ also 
becomes constant. Then it follows that $A$ also must be constant, and hence 
$\theta=0$ along $H$. This immediately yields 
\bea 
R_{\mu\nu}l^\mu l^\nu=-\sigma_{ij}\sigma^{ij} \,. 
\eea 
Since the null energy condition is imposed, it must be that 
$R_{\mu\nu}l^\mu l^\nu \ge 0$. This would imply  
that $2\sigma_{ij} = {\cal L}_l g_{ij}=0$ for any $i$ and $j$. 
By using $l=fn$ and $g_4(\p_i,\,n)=0$, we have  
\bea 
{\cal L}_l g_{ij}=f{\cal L}_{n} g_{ij}=f{\cal L}_\eta g_{ij}=0 \,.  
\eea
Therefore $\eta$ must be a Killing vector on $\Sigma$. 
\hfill $\Box$ 

\medskip 

It is worth mentioning to the other Bianchi models. In general, $\theta\neq 0$, as $\det {\Lambda^I}_i$ depends on the 
local coordinate $x^i$. 
In this case, $\theta$ must be positive in the future direction 
according to the area theorem or the second law of the thermodynamics, implying 
that the entropy density continues to increase. 
As suggested in the inhomogeneous models \cite{Figueras:2012rb, Fischetti:2012vt}, such a dissipating stationary black brane 
solution could exist even in the homogeneous case.

\section{Symmetries in the bulk: A no-go in Einstein-Maxwell-dilaton theory}
\label{section3}
In this section, we will complete our proof that momentum must be zero 
along the directions with no translational symmetry 
by showing that the shift vector $\tilde{N}_I$ in Eq.~(\ref{metric_tr}) 
is identically zero outside the horizon. 
As we mentioned in Sec.~\ref{sec:intro}, the proof crucially depends on 
whether gauge field acquires a mass or not. 
In fact, if $U(1)$ gauge field acquires the mass, the gauge field outside the horizon can possess a momentum along the 
direction with no translational symmetry~\cite{Iizuka2013Dec}.

Since we are interested in the case where the mass of the gauge field is zero, 
we proceed by restricting ourselves to the following 
Einstein-Maxwell~(dilaton) theory: 
\begin{align}
\label{action-EMD}
S=\int d^5x \sqrt{-g}\left(R-(\nabla\phi)^2-\frac{1}{4}e^{2\alpha \phi}F^2-\frac{1}{4}W^2-V(\phi)\right), 
\end{align}
where $A_\mu$ and $B_\mu$ are gauge potential one-forms 
and their field strengths are $F=dA$ and $W=dB$, respectively.  
Here, $\alpha$ is an arbitrary parameter and $V(\phi)$ is an arbitrary potential of the dilaton field $\phi$ such that $V$ has 
an extreme value $-12/L^2$ at $\phi=\phi_0$ and satisfies the BF bound. The one-form $B$ plays a role for inducing 
the helical lattice effect on the bulk in Bianchi VII$_0$ black brane spacetime, as shown in \cite{Donos:2012js, Iizuka2013Dec}.    

The equations of motion are 
\bea
\label{Eq:Ein_EMD}
&& R_{\mu\nu}=\nabla_\mu\phi \nabla_\nu\phi+\frac{1}{2}e^{2\alpha\phi}F_{\mu\rho}{F_\nu}^\rho
+\frac{1}{2}W_{\mu\rho}{W_\nu}^\rho \quad \quad \quad \nonumber \\
&& \qquad \qquad -g_{\mu\nu}\left(\frac{1}{12}(e^{2\alpha\phi}\,F^2+W^2)-\frac{1}{3}V(\phi)  \right) \,, \\
\label{Eq:Maxwell_EMD}
&& \sqrt{-g}\,\nabla_\nu(e^{2\alpha\phi}F^{\mu\nu})=\p_\nu(\sqrt{-g}e^{2\alpha\phi}F^{\mu\nu})=0 \,, \\ 
\label{Eq:Maxwell_B}
&& \sqrt{-g}\,\nabla_\nu W^{\mu\nu}=\p_\nu(\sqrt{-g}W^{\mu\nu})=0 \,, \\ 
\label{Eq:dilaton_EMD}
&& \nabla^2\phi-\frac{1}{4}\alpha e^{2\alpha\phi}F^2-\frac{1}{2}\frac{\p V}{\p \phi}=0 \,.  
\eea

To simplify the analysis, we shall restrict the metric~(\ref{metric_tr}) into 
the following particular form 
\bea 
\label{metric_diagonal}
ds^2=-fdt^2+\frac{dr^2}{f}+e^{2v_3}(dx^1+\Omega\,dt)^2+e^{2v_1}(\omega^1)^2+e^{2v_2}(\omega^2)^2, 
\eea 
where $\p_{x^1}$ is a direction of no translational symmetry, as explained below. 
We assume $f, \Omega, v_i$ are functions of $r$ only. 
In general, the asymptotic behavior of the functions $f$ and $v_i$ are given by 
\bea 
\label{asy_behavior}
f=\frac{r^2}{L^2}+O(1), \qquad v_i\simeq \ln r+\frac{1}{2}\ln C_i, \qquad \Omega=\Omega_0+\frac{\Omega_1}{r^4},  
\eea 
where $\Omega_0$, $\Omega_1$ are constants and $C_i$ are positive constants. 
Since we are seeking for a solution where momentum flow occurs 
without non-normalizable mode, we shall impose 
\bea 
\label{asy_cond_Omega}
\Omega_0=0
\eea 
as an asymptotic condition for $\Omega$. 

Note that the subleading constant term in $f$ comes from the curvature of the three dimensional homogeneous Bianchi space. 
So, strictly speaking, the geometry does not asymptotically approach 
AdS spacetime, which is conformally flat. However, in this paper, let us 
simply call such a geometry by asymptotically ``AdS''spacetime
in the sense that the leading terms in the metric approach AdS
spacetime. In the following subsections, we will show that $\Omega$ is identically zero in Bianchi II, VI$_0$, and VII$_0$.

\subsection{Bianchi VII$_0$ model}
Let us start with most interesting case, Bianchi type VII$_0$. 
Bianchi type VII$_0$ is defined in (\ref{BianchitypeVII0}), which has three Killing vectors,  
\bea
\xi_1 = \p_{x^2} \,, \quad \xi_2 = \p_{x^3} \,, \quad \xi_3=\p_{x^1}-x^3 \p_{x^2}+x^2 \p_{x^3} \,,
\eea
and three invariant one-forms, 
\bea
\omega^1=\cos(x^1)dx^2+\sin(x^1)dx^3 , \, \omega^2=-\sin(x^1)dx^2+\cos(x^1)dx^3 , \,  \omega^3=dx^1 \,.
\eea
From these, it is clear that along $x^1$, there is no translational 
invariance, $x^1 \to x^1 + const$. 

Following~\cite{Iizuka2013Dec}, we make the following ansatz 
for the Maxwell field, the vector field $B$, and the dilaton field $\phi$,  
\bea
A_\mu dx^\mu=A_t(r)dt +A_{x^1}(r) \omega^3 \,, \qquad 
B_\mu dx^\mu=b(r)\omega^1, \qquad \phi=\phi(r).  
\eea
Under the ansatz, we obtain equations of motion for $A_t$ and $A_{x^1}$ as 
\bea
\label{eq:t_Maxwell_EMD}
\p_r(e^{v_1+v_2+v_3+2\alpha\phi}(A_t'-\Omega A_{x^1}')) &=&0 \,,\\
\label{eq:x_Maxwell_EMD}
\p_r(e^{v_1+v_2+v_3+2\alpha\phi}\{\Omega(A_t'-\Omega A_{x^1}')+fe^{-2v_3}A_{x^1}'\})&=&0 \,,   
\eea
where here and hereafter the {\it prime} denotes 
the derivative with respect to $r$. These equations are integrated to 
\bea
\label{sol:gauge1_EMD}
e^{v_1+v_2+v_3+2\alpha\phi}(A_t'-\Omega A_{x^1}') &=&\beta_0 \,, \\
\label{sol:gauge2_EMD}
e^{v_1+v_2+v_3+2\alpha\phi}\{\Omega(A_t'-\Omega A_{x^1}')+fe^{-2v_3}A_{x^1}'\}&=&C \,,   
\eea
where $\beta_0$ and $C$ are integration constants. 
By the asymptotic conditions (\ref{asy_behavior}), (\ref{asy_cond_Omega}), and $\phi\to \phi_0$, 
we obtain $A_t'(r\to \infty)\sim \beta_0/r^3$. So, $\beta_0$ corresponds to the expectation value of 
the charge density, $\Exp{j_t}$ in the dual field theory by the AdS/CFT duality. Since we are interested in 
the dual field theory where charge density or charge current exists, hereafter, we assume that $\beta_0\neq 0$.   

Eliminating $A_t$ from these two equations~(\ref{sol:gauge1_EMD}), (\ref{sol:gauge2_EMD}), one obtains 
\begin{align}
\label{sol:gauge3_EMD}
\Omega \beta_0+fe^{2\alpha\phi}e^{v_1+v_2-v_3}A_{x^1}'=C. 
\end{align} 
Below we discuss the two cases, the one $v_1 \neq v_2$ and the other 
$v_1=v_2$, on the horizon, separately. 

Now, first consider the generic case, $v_1\neq v_2$ on the horizon. 
Then, $\p_{x^1}$ is the direction of no translational invariance 
and hence $\Omega_h=0$ by the theorem in Sec.~\ref{section2}. 
This implies that $C=0$, as $f=0$ on $H$. 
%
%
%
%
The equation of motion for $\Omega$ is obtained from the $(t, x^1)$ and $(x^1, x^1)$ component 
of the Einstein equation (\ref{Eq:Ein_EMD}) by eliminating $v_3''$, as  
\begin{align}
\label{Eq:Omega_EMD}
& f\Omega''+f(v_1'+v_2'+3v_3')\Omega' \nonumber \\
&- e^{-2 (v_2 + v_3)} \left[ 
   e^{2 v_2} \left(4 \Omega \sinh^2\left( v_1 - v_2 \right) -  e^{2 \alpha \phi}   f A_{x^1}' ( A_t' - \Omega A_{x^1}' ) \right) + b^2 \Omega \right]=0. 
\end{align}
Eliminating $A'_{x^1}$ and $A'_t$ from Eqs.~(\ref{sol:gauge1_EMD}) and (\ref{sol:gauge3_EMD}) under the condition $C=0$, 
we obtain a linear differential equation with respect to $\Omega$, 
\begin{align}
\label{eq_Omega_linear}
\Omega''+(v_1'+v_2'+3v_3')\Omega'-\frac{e^{-2v_3}}{f}[  \beta_0^2e^{-2(\alpha \phi+v_1+v_2)}+e^{-2v_2}b^2+4\sinh^2(v_1-v_2)]\Omega=0. 
\end{align}  

Now suppose there were a nontrivial solution of this equation that 
satisfies the two boundary conditions $\Omega_h = 0$ 
and $\Omega(\infty) = \Omega_0=0$. This would imply that 
$\Omega(r)$ must admit a maximal (or minimal) value somewhere 
if $\Omega > 0$ (or $<0$, respectively). Having a maximal (minimal) 
value implies that at some radius, say $r = r_{ex}$, $ \Omega'(r_{ex})=0$, 
and $ \Omega''(r_{ex}) < 0$ ($>0$), so that it admits local Maximal 
(Minimal), and thus that 
\bea
\Omega(r_{ex}) \Omega''(r_{ex}) < 0\,.
\eea
However this contradicts with (\ref{eq_Omega_linear}), since 
\bea
\frac{e^{-2v_3}}{f}[  \beta_0^2e^{-2(\alpha\phi+v_1+v_2)}+e^{-2v_2}b^2+4\sinh^2(v_1-v_2)] > 0 \,. 
\eea
Therefore it is impossible to obtain a solution $\Omega$ which is 
not identically zero; the only allowed solution is $\Omega = 0$ 
in all the radius.

Next, we consider the case $v_1 = v_2$ on the horizon, for which the 
situation is a little bit complicated. 
In this case, the translational symmetry along $x^1$ direction is 
recovered on the horizon because the three dimensional spatial 
metric becomes flat space under our metric ansatz~(\ref{metric_diagonal}). 
Then, the theorem in Sec.~\ref{section2} cannot exclude the possibility that 
the horizon rotates along $x^1$ direction. 

Let us investigate the equations for $\xi:= v_1-v_2$ and $b$,  
\bea
\label{eq_xi_SC}
& f\xi''+\{f'+f(v_1'+v_2'+v_3')\}\xi'-2\left(e^{-2v_3}-\frac{\Omega^2}{f}\right)\sinh 2\xi \nonumber \\
& \qquad \qquad \qquad \quad -\frac{e^{-2(v_2+v_3)}}{2}\left(1-e^{2v_3}\frac{\Omega^2}{f} \right)b^2
+\frac{1}{2}e^{-2v_1}fb'^2=0 \,, \\ 
\label{Eq_b_SC}
& fb''+\{f'+f(v_3'+v_2'-v_1') \}b'-e^{2(v_1-v_2)}\left(e^{-2v_3}-\frac{\Omega^2}{f}\right)b=0. 
\eea
As shown in \cite{Iizuka2013Dec}, the near horizon behavior of $b$ is 
\bea 
\label{singular-b}
b\sim (r-r_h)^{\eta_\pm}, \qquad \eta_\pm=\pm \frac{i\Omega_h}{\kappa}, 
\eea 
where $\kappa=f'(r_h)$. Since both the solutions are singular at the horizon, we conclude that their coefficients 
must vanish in order to have a smooth horizon. This means that $b(r)=b'(r)=0$ 
near $r\to r_h$. Since 
we solve the 2nd order differential equation, this means that  $b(r)=0$ in all the radius. Substituting $b=0$ into 
Eq.~(\ref{eq_xi_SC}), we obtain similar singular solutions near the horizon as 
\bea 
\label{xi_singular_behavior}
\xi\sim (r-r_h)^{2\eta_\pm},
\eea
where we used the fact that $\xi=0$ on the horizon. Thus, to have a smooth horizon, $\Omega_h=0$. Repeating the 
same argument on Eq.~(\ref{eq_Omega_linear}), we obtain the same conclusion 
that $\Omega=0$ at all the radius outside the horizon.  

Given $\Omega =0$, 
it is easy to see that the expectation value of the current on the dual field theory, $\Exp{j_{x^1}}$ along $x^1$ direction must be 
zero. By Eq.~(\ref{sol:gauge3_EMD}), we obtain $A_{x^1}'=0$. Since $\Exp{j_{x^1}}$ corresponds to the subleading term 
in the asymptotic behavior of $A_{x^1}$, 
\begin{align}
\label{asy_sol_Ax}
A_{x^1}\simeq A_1+\frac{A_2}{r^2},  
\end{align}
$A_{x^1}'=0$ means $A_2=0$. Thus, we obtain $\Exp{j_{x^1}}=0$, implying that the charge current is zero along $x^1$ direction, as 
expected.  

\subsection{Bianchi VI$_0$}
Next, in this subsection, we consider the Bianchi VI$_0$ case. 
The corresponding three Killing vectors $\xi_i~(i=1,2,3)$ are 
\begin{align}
\label{VI_0_killing}
\xi_1=\p_{x^2}, \qquad \xi_2=\p_{x^3}, \qquad \xi_3=\p_{x^1}+ x^2 \p_{x^2}- x^3 \p_{x^3} \,,
\end{align}
and the invariant one-forms are 
\begin{align}
\label{VI_h_vectorbasis}
& \omega^1=e^{-x^1}d x^2, \qquad \omega^2=e^{x^1}d x^3, \qquad \omega^3=dx^1. 
\end{align}
Clearly there is no translational invariance along $x^1$.  
Then we obtain $\Omega_h=0$ at the horizon by the theorem in Sec.~\ref{section2}. 

From the Maxwell equation (\ref{Eq:Maxwell_EMD}),  
\begin{align}
\label{VI_h_gauge}
& e^{v_1+v_2+v_3+2\alpha \phi}(A_t'-\Omega A_{x^1}')=\beta_0, \nonumber \\
& e^{v_1+v_2+v_3+2\alpha \phi}\{\Omega(A_t'-\Omega A_{x^1}') + fe^{-2v_3}A_{x^1}'\}=C. 
\end{align}
Since $\Omega_h=0$, we obtain $C=0$. Thus, similar to the Bianchi VII$_0$ case, 
we obtain a linear differential equation for $\Omega$, 
 \begin{align}
\label{VI_h_Omega1}
f\Omega''+f(v_1'+v_2'+3v_3')\Omega'-e^{-2v_3}[4+e^{-2v_1}b^2+\beta_0^2e^{-2(\alpha\phi+v_1+v_2)}]\Omega=0 
\end{align}
from the Einstein equation. 
Just like the Bianchi VII$_0$ case, we have 
\bea
e^{-2v_3}[4+e^{-2v_1}b^2+\beta_0^2e^{-2(v_1+v_2)}] > 0. 
\eea 
This implies that $\Omega$ cannot have a solution satisfying $\Omega(r_h) = \Omega(\infty) = 0$, and thus, 
$\Omega = 0$ in all the radius.   

\subsection{Bianchi II}
In the case of the Bianchi II, the three Killing vectors are 
\begin{align}
\label{II_killing}
\xi_1=\p_{x^2}, \qquad \xi_2=\p_{x^3}, \qquad \xi_3=\p_{x^1}+x^3 \p_{x^3}
\end{align}
and the invariance one-forms are 
\begin{align}
\label{VI_h_vectorbasis}
& \omega^1=dx^2 -x ^1dx^3, \qquad \omega^2=dx^3, \qquad \omega^3=dx^1. 
\end{align}
 
Clearly there is no translational invariance along $x^1$.  
Then we obtain $\Omega_h=0$ at the horizon by the theorem in Sec.~II.  

The Maxwell equation (\ref{Eq:Maxwell_EMD}) gives the same form as the Bianchi VII$_0$ case.  
Similarly,  
we obtain a linear differential equation for $\Omega$   
\begin{align}
\label{II_Omega1}
f\Omega''+f(v_1'+v_2'+3v_3')\Omega'-e^{-2v_3}[e^{2(v_1-v_2)}+e^{-2v_2}b^2
+\beta_0^2e^{-2(\alpha\phi+v_1+v_2)} ]\Omega= 0 \,.
\end{align}
This equation denies the solution where $\Omega$ has local 
maxima/minima so that the boundary conditions 
$\Omega(r_h) = \Omega(\infty) = 0$ are satisfied. 
  
We have shown that $\Omega$ is identically zero outside the horizon in Bianchi II, VI$_0$, and 
VII$_0$ cases by using the assumption $\beta_0\neq 0$. 
Note that this assumption is not necessarily required in Bianchi II, VI$_0$ 
cases because we can obtain $\Omega(r_{ex}) \Omega''(r_{ex}) < 0$ regardless 
of $\beta_0\neq 0$. In Bianchi VII$_0$ case, however, we need this assumption 
to prove $\Omega(r_{ex}) \Omega''(r_{ex}) < 0$. So, it is interesting whether 
we can show that $\Omega$ is identically zero outside the horizon 
for the Bianchi VII$_0$ case even when $\beta_0=0$, or the vacuum case. 
We leave this question open for future work. 
  
\section{Symmetries broken in the bulk, restored on the horizon} 
\label{section4}

So far we have seen that it is difficult to have a gravitational solution on 
the Bianchi type II, VI$_0$, VII$_0$, where $\Omega$ is nonzero 
along a direction on which there is no translational invariance. 
One possible exception is that 
the gauge potential $A$ in Sec.~\ref{section3} has a mass, just like the case 
in which a gauge field acquires a mass according to 
the superconducting state~\cite{Iizuka2013Dec}. 
However in that case, the black brane horizon itself is actually not rotating and momentum is carried only by the matter fields outside. 
Therefore, it would be interesting to explore whether there exists 
a stationary black brane solution, where the horizon is rotating and 
the momentum flows along the direction with no translational symmetry, 
evading the symmetry theorem in Sec.~\ref{section2}. 
As explained in Sec.~\ref{section3}, this is only possible if $v_1=v_2$ on the horizon in the Bianchi type VII$_0$ because 
translational symmetry is recovered on $\p_{x^1}$ direction on the horizon. 

In this section, we will construct such a solution 
in the Einstein-Maxwell theory with an additional vector field $C$ that has 
a potential with a source term. 
The horizon is rotating along $x^1$ direction on which translational symmetry is broken outside the horizon. 
Such a solution is different from the rotating stationary solutions numerically found in \cite{Figueras:2012rb, Fischetti:2012vt} 
in the sense that no dissipation occurs in our solution.

\subsection{Our model} 
In the Bianchi VII$_0$, the model we consider is 
\bea
\label{action_two_vector} 
 S &=& \int d^5x \sqrt{-g}\left(R+\frac{12}{L^2}-\frac{1}{4}F^2-\frac{1}{4}W^2-V(C) \right),  \,\,\,\,  
\eea
where $A$ and $C$ are one-form potentials, 
\bea
 F=dA \,,\quad W=dC \,,\quad 
\eea
and the potential term $V(C)$ for the one-form $C$ is given by 
\bea
V(C) &=& a_0 (C - C_0)^2+a_1 (C - C_0)^4 \,, \\
C_0  & \equiv &  c_0 \, \omega^1 \,,
\eea
where $\omega^1$ is the invariant one-form. 
Note that the gauge symmetry for $C$ is explicitly broken by 
their potential $V(C)$. We regard here, $a_0$, $a_1$, $c_0$ as parameters 
inducing the lattice effects. Especially $c_0 \neq 0$ is crucial 
for the {\it source term} to be introduced.  
Note also that the gauge field $A$ corresponds to normal states in 
the dual field theory and accordingly does not have a mass term 
in (\ref{action_two_vector}).

This Lagrangian does not have a translational invariant vacuum 
if these parameters are nonzero, especially $c_0 \neq 0$.  
We regard that this is a sort of effective Lagrangian for the model where the lattice effects are introduced so that 
we do not have a translationally invariant vacuum. How translational invariant UV theory 
could induce such a source term by symmetry breaking is another question. 
In this section, putting aside such a question, 
we regard the action (\ref{action_two_vector}) 
as given and seek for a solution under the metric 
ansatz (\ref{metric_diagonal}).

We assume that the flux forms 
\begin{align}
C_\mu dx^\mu=c(r)\omega^1, \qquad 
\end{align}
in such a way that at the horizon, 
\bea
\lim_{r \to r_h} c(r) \to 0 \,.
\eea 
It is rather easy to see that without the potential $V(C)$, 
the only allowed regular solution is $v_1 = v_2$ in all the radius, 
implying that the lattice effects disappear. 
To see this explicitly, let us first consider the case where $a_0=a_1=0$. 
Then, the equations of motion for $\xi=v_1-v_2$ and $c$ are equivalent to 
Eqs.~(\ref{eq_xi_SC}) and (\ref{Eq_b_SC}) by replacing $b$ with $c$. So, following 
the argument in Sec.~\ref{section3}, we obtain $\xi=c=0$ in all the radius when $a_0=a_1=0$. 

{\subsection{Equations of motion}}
We now consider generic $a_0\neq 0$, $a_1\neq 0$ case for the action (\ref{action_two_vector}). 
The Einstein equation is  
\begin{align}
\label{Einstein_two_vector}
& R_{\mu\nu}=\frac{1}{2}F_{\mu\alpha}{F_\nu}^\alpha +\frac{1}{2}W_{\mu\alpha}{W_\nu}^\alpha
+\{a_0+2a_1 (C-C_0)^2\}(C-C_0)_\mu (C-C_0)_\nu 
\nonumber \\
&\qquad \qquad -\frac{g_{\mu\nu}}{12}\left[\frac{48}{L^2}+F^2+W^2+4 a_1(C-C_0)^4  \right]. 
\end{align}
Using the metric ansatz (\ref{metric_diagonal}), 
it is straightforward to obtain the equation of motion for $\xi=v_1-v_2$ as 
\bea
\label{eq_xi_two_vector}
&& f\xi''+\{f'+f(v_1'+v_2'+v_3')\}\xi' 
- \, 2\left(e^{-2v_3}-f^{-1}{\Omega^2}\right)\sinh 2\xi \nonumber \\
&& \qquad  - \, \frac{e^{-2(v_2+v_3)}}{2}\left(1-e^{2v_3}f^{-1} \Omega^2 \right)c^2
+\frac{1}{2}e^{-2v_1}fc'^2 \nonumber \\
&& \qquad =-\, a_0e^{-2v_1}(c-c_0)^2-2a_1e^{-4v_1}(c-c_0)^4 \,. \quad 
\eea
The equations of motion for $\Omega(r)$, $v_1$, $v_3$ are respectively, 
\bea
\label{eq_rotating_Omega1}
&& \Omega''+(v_1'+v_2'+3v_3')\Omega'-4 {e^{-2v_3}} f^{-1} \sinh^2(v_1-v_2)\Omega \nonumber \\
&& \qquad + \, e^{-2v_3}A_x'(A_t'-\Omega A_x')
-\frac{e^{-2(v_2+v_3)}}{f}c^2\Omega=0 \,, \\
\label{eq_rotating_v1_1}
&& fv_1''+\{f'+f(v_1'+v_2'+v_3')\}v_1'+\frac{1}{6}A_t'^2 
+\, \frac{1}{6}(\Omega^2-fe^{-2v_3})A_x'^2
 -\frac{1}{3}\Omega A_x'A_t' 
  \nonumber \\
&& \qquad 
+\sinh\{2(v_1-v_2) \}f^{-1} \Omega^2
-e^{-2v_3}\sinh\{2(v_1-v_2)\} 
-\frac{4}{L^2}+a_0e^{-2v_1}(c-c_0)^2 \nonumber \\
&& \qquad 
+\frac{5}{3}a_1e^{-4v_1}(c-c_0)^4
+\frac{1}{3}e^{-2v_1}fc'^2
-\frac{e^{-2(v_2+v_3)}c^2}{6}\left(1-{e^{2v_3}}{f^{-1}}\Omega^2  \right)
 =0 \,, \\
%
\label{eq_rotating_v3_1}
&& fv_3''+\{f'+f(v_1'+v_2'+v_3')\}v_3'+\frac{1}{6}(A_t'-\Omega A_x')^2 
+\frac{1}{3}fe^{-2v_3}A_x'^2
\nonumber \\
&& \qquad
 +2e^{-2v_3}\sinh^2(v_1-v_2)+\frac{e^{2v_3}}{2}\Omega'^2
 -\frac{4}{L^2}
-\frac{a_1}{3}e^{-4v_1}(c-c_0)^4
\nonumber \\
&& \qquad 
-\frac{1}{6}fe^{-2v_1}c'^2
+\frac{c^2}{6}e^{-2(v_2+v_3)}\left(2+{e^{2v_3}} f^{-1}\Omega^2  \right) =0 \,. 
\eea
From the constraint equation, we have for $f$ as 
\begin{align}
\label{eq_rotating_Constr1}
& \frac{1}{2}e^{2v_3}\Omega'^2-\frac{1}{2}fe^{-2v_3}A_x'^2+\frac{1}{2}(A_t'-\Omega A_x')^2
-\frac{12}{L^2}
 +2\sinh^2(v_1-v_2)\left(e^{-2v_3}
  - f^{-1} \Omega^2  \right)
  \nonumber \\
&\qquad
+f'(v_1'+v_2'+v_3')  +2f(v_1'v_2'+v_2'v_3'+v_3'v_1') +a_0e^{-2v_1}(c-c_0)^2 \nonumber \\
&\qquad +a_1e^{-4v_1}(c-c_0)^4
-\frac{f}{2}e^{-2v_1}c'^2
+\frac{e^{-2(v_2+v_3)}}{2}\left(1-e^{2v_3}f^{-1} \Omega^2\right)c^2=0 \,. 
\end{align}
The equations of motion for the potential $A$ give
\bea
e^{v_1+v_2+v_3}(A_t'-\Omega A_x') &=&\beta_0 \,, \\
\label{sol:gauge2_1}
\beta_0\Omega+e^{v_1+v_2-v_3}fA_x'&=&\beta_1  \,, 
\eea
and for $C$, 
\bea
\label{eq:c_two-vector}
&& fc''+\{f'+f(v_3'+v_2'-v_1') \}c' 
-e^{2(v_1-v_2)}\left(e^{-2v_3}- f^{-1}{\Omega^2}\right)c
\nonumber \\
&&\qquad 
 =2a_0(c-c_0)+4a_1 e^{-2v_1}(c-c_0)^3  \,. 
\label{sol:gauge1_1}
\eea
Now, we have additional terms due to $a_0$ and $a_1$, and {Eq.~(\ref{eq:c_two-vector}) is 
a non-linear differential equation with respect to $c$}. {Note that $c$ must vanish on the horizon 
to have a regular solution for $c$, otherwise the l.~h.~s. of Eq.~(\ref{eq:c_two-vector}) would diverge at the 
horizon, due to the term $f^{-1}{\Omega^2}c$ when $\Omega_h\neq 0$. 
This implies that the r.~h.~s. of Eq.~(\ref{eq_xi_two_vector}) becomes non-vanishing on the horizon}. 
By the theorem in Sec.~\ref{section2}, we must set $\xi=0$ on the horizon. 
Because of these, we can now expand $c$, $v_i$, and $\Omega$ as 
\bea
\label{expansion}
c(r) &=& c_1 ( r - r_h )  + O(r - r_h)^2 \,, \nonumber \\
v_ 1 &=& v_2 + O(r - r_h) \,, \nonumber \\
\Omega(r) &=& \Omega_h+O(r-r_h) \,, \nonumber \\
c_1 &=& - \frac{2 \kappa c_0}{\kappa^2 + \Omega_h^2} \left( a_0 + 2 a_1 e^{-2 v_1} c_0^2\right)
\eea
to have a regular solution, here the coefficient $c_1$ is determined by Eq.~(\ref{eq:c_two-vector}).  
On the other hand, at spatial infinity $r \to \infty$, $c(r)$ behaves as  
\bea
c(r)\simeq \alpha_+r^{n_+}+\alpha_-r^{n_-}+\left(1-\frac{e^{2\xi(\infty)}}{2C_3a_0}r^{-2}\right)c_0, \quad 
n_\pm=-1\mp\sqrt{1+2a_0L^2}. 
\eea
According to the AdS/CFT dictionary, the non-normalizable mode $r^{n_-}$ and the normalizable mode $r^{n_+}$ 
correspond to the source term and the expectation value on the field theory side, 
respectively. So, we impose a boundary condition,  
\bea 
\label{num_con1}
\alpha_-=0. 
\eea
In addition, we also impose that 
\bea
\label{num_con2}
\Omega(r=\infty)=0, 
\eea 
since we are interested in the boundary theory on a static spacetime.   

There still remains a gauge freedom for the choice of the coordinates $t$, $r$, $x^2$, and $x^3$ 
under the boundary conditions~(\ref{num_con1}) and (\ref{num_con2}), due to the scaling freedom  
\bea 
\label{free_parameters}
&&r\to \alpha_1 r, \quad t\to \frac{t}{\alpha_1}, \quad x^2 \to \alpha_2\, x^2, \quad  x^3 \to \alpha_2\, x^3,   \\
&& f \to \alpha^2_1 f, \quad e^{v_1} \to \frac{e^{v_1}}{\alpha_2}, \quad e^{v_2} \to \frac{e^{v_2}}{\alpha_2},
\eea
which preserves the metric ansatz (\ref{metric_diagonal}) and the asymptotic AdS boundary condition~(\ref{asy_behavior}) with $\Omega_0=0$. 
Thus, to perform the numerics and determine the temperature uniquely, we shall fix the asymptotic AdS metric as 
\bea 
\label{bd_AdS_scaling}
ds^2\simeq -\frac{r^2}{L^2}dt^2+\frac{L^2}{r^2}dr^2+r^2(dx^1+\Omega \,dt)^2+r^2e^{\xi}(\omega^1)^2+r^2e^{-\xi}(\omega^2)^2.
\eea
\subsection{Numerical solution}
Under the boundary conditions~(\ref{num_con1}) and (\ref{num_con2}), we numerically construct a rotating 
black brane solution in asymptotically AdS spacetime.  
By imposing regularity on the horizon, one can derive the following boundary conditions;   
\bea
\label{initial_con_xi}
&& \xi'(r_h)=-\frac{\kappa\, b_0^2}{\kappa^2+4\Omega_h^2}(a_0+2a_1e^{-2v_1(r_h)}b_0^2)e^{-2v_1(r_h)} \,, \\
\label{initial_con_v1}
&& \kappa v_1'(r_h)+\frac{\beta_0^2}{6}e^{-2(v_1(r_h)+v_2(r_h)+v_3(r_h))}+\frac{2\xi'(r_h)\Omega_h^2}{\kappa}
-\frac{4}{L^2} \nonumber \\
&&\qquad \qquad +a_0e^{-2v_1(r_h)}b_0^2+\frac{5}{3}a_1e^{-4v_1(r_h)}b_0^4=0 \,,\\ 
\label{initial_con_v3}
&& \kappa v_3'(r_h)+\frac{\beta_0^2}{6}e^{-2(v_1(r_h)+v_2(r_h)+v_3(r_h))} \nonumber \\
&& \qquad \qquad +\frac{e^{2v_3(r_h)}}{2}\Omega'^2(r_h)-\frac{4}{L^2}-\frac{a_1}{3}e^{-4v_1(r_h)}b_0^4=0 \,, \\
\label{initial_cond_gauge}
&& \beta_1=\beta_0\Omega_h \,, 
\eea
where the last equation is derived from Eq.~(\ref{sol:gauge2_1}). 
Thus, the free parameters on the horizon are 
\bea
\label{initial_parameters} 
r_h, \,\, v_1(r_h),\,\, v_3(r_h),\,\, \Omega_h,\,\, \Omega'(r_h), \,\, \beta_0, \,\, \kappa. 
\eea
{Note that 
the solution for $c$ is uniquely determined by the regularity condition at 
the horizon~(\ref{expansion}) for the given parameters~(\ref{initial_parameters}), and the asymptotic behavior does not 
generically match on to the condition~(\ref{num_con1}). 
So, to match on to the UV boundary condition~(\ref{num_con1}), we need to 
choose a specific value $a_1$ for each parameter choice~(\ref{initial_parameters}). In addition, to satisfy the UV boundary 
condition~(\ref{num_con2}), we will find a specific parameter $\Omega'(r_h)$ for the given values 
$v_1(r_h)$, $v_3(r_h)$, $\Omega_h$, $\beta_0$, $\kappa$ by a shooting method~\footnote{Here, $r_h$ has gauge degrees of freedom, and we fix it 
by imposing $f$ to approach $r^2/L^2+O(1)$ asymptotically.}. 
This implies that we need to choose a theory $a_0$, $a_1$, and $c_0$ 
for a given temperature $T=\frac{\kappa}{4\pi}$, angular velocity $\Omega(r_h)$, and the charge density $\beta_0$.}  

Let us pay attention to the case $a_0>0$, otherwise, the null energy condition would be violated asymptotically, as seen 
below. Then, if the condition (\ref{num_con1}) is satisfied, $c$ asymptotically 
approaches $c_0$. We find two solutions by a shooting method and {the parameter choices are 
\bea 
\mbox{(I)}:\,\,\,& L=\sqrt{2},\,\, 
  v_1(r_h)=-0.1161, \,\,v_3(r_h)=0, \,\,a_0=0.1, \,\,a_1=-10.82 \nonumber \\
& \Omega(r_h)=0.3120, \,\,\Omega'(r_h)=-1.213, \,\,\beta_0=0.4861,\,\, \kappa=1.560,  
\eea 
\bea 
\mbox{(II)}: \,\,\, & L=\sqrt{2},\,\,v_1(r_h)=-0.04137, \,\,v_3(r_h)=0, \,\,a_0=0.1, \,\,a_1=-11.17, \nonumber \\
& \Omega(r_h)=0.1808, \,\,\Omega'(r_h)=-0.7043, \,\,\beta_0=0.5644, \,\,\kappa=1.811,
\eea
as shown in Figure 1 - 8.}
\begin{figure}[htbp]
 \begin{minipage}{0.5\hsize}
  \begin{center}
   \includegraphics[width=70mm]{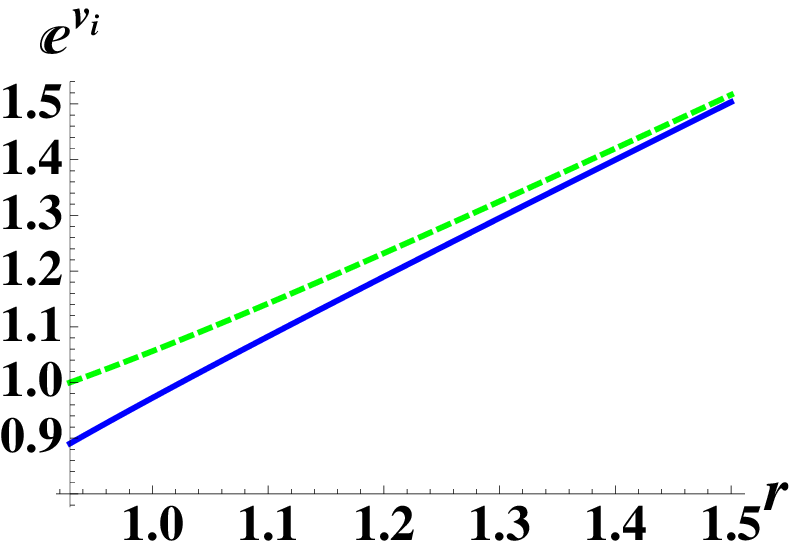}
  \end{center}
  \caption{$e^{v_1}$~(solid) and $e^{v_3}$~(dashed) are shown for the parameter choice (I). }
  \label{fig:one}
 \end{minipage}
 \begin{minipage}{0.5\hsize}
  \begin{center}
   \includegraphics[width=70mm]{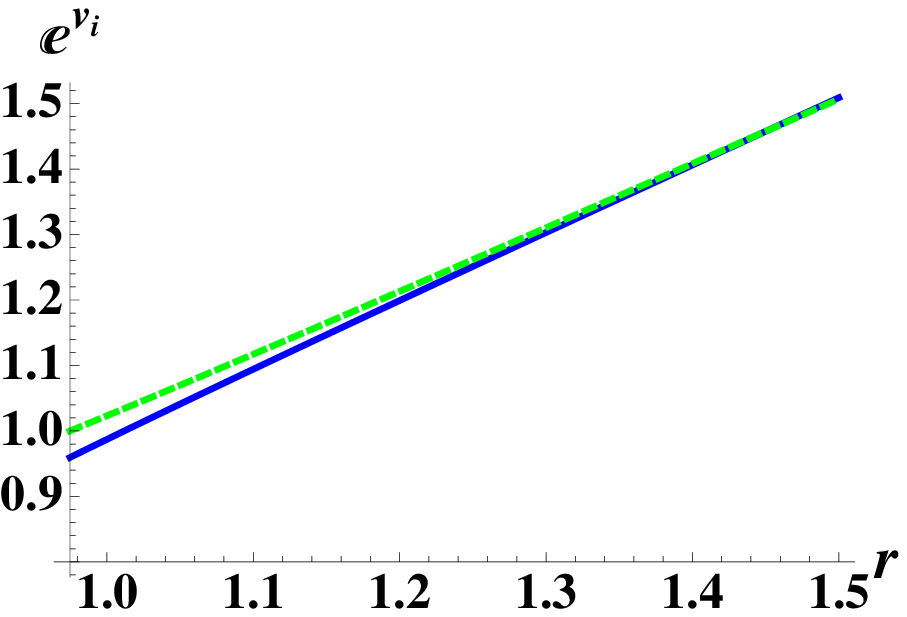}
  \end{center}
  \caption{$e^{v_1}$~(solid) and $e^{v_3}$~(dashed) are shown for the parameter choice (II).}
  \label{fig:two}
 \end{minipage}
\end{figure}
\begin{figure}[htbp]
 \begin{minipage}{0.5\hsize}
  \begin{center}
   \includegraphics[width=70mm]{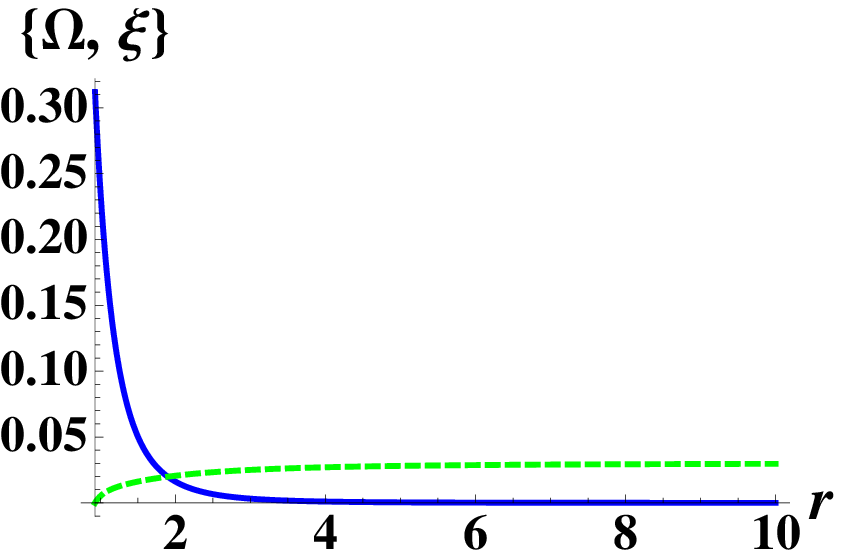}
  \end{center}
  \caption{$\Gamma=\Omega$~(solid) and $\Gamma=\xi$~(dashed) are shown for the parameter choice (I). }
  \label{fig:one}
 \end{minipage}
 \begin{minipage}{0.5\hsize}
  \begin{center}
   \includegraphics[width=70mm]{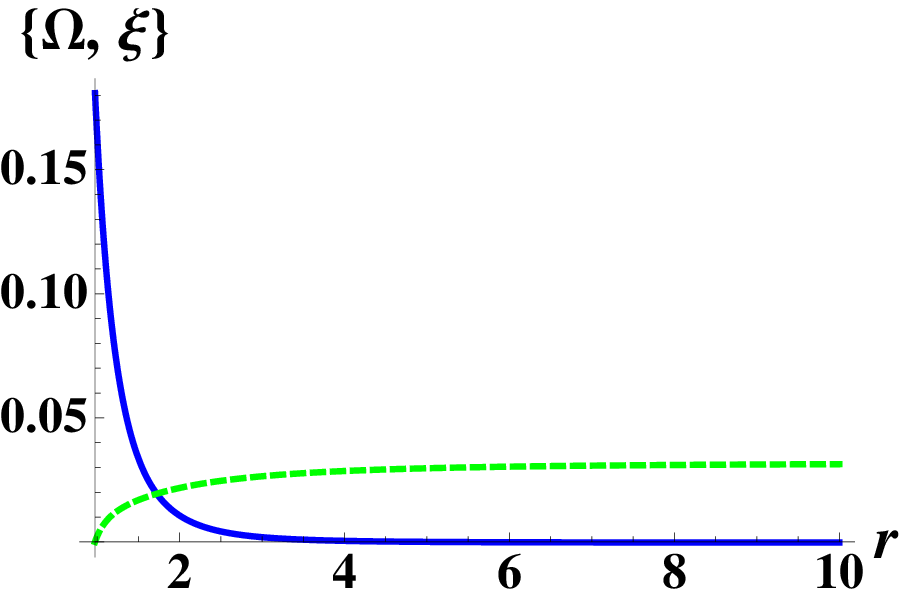}
  \end{center}
  \caption{$\Gamma=\Omega$~(solid) and $\Gamma=\xi$~(dashed) are shown for the parameter choice (II).}
  \label{fig:two}
 \end{minipage}
\end{figure}
\begin{figure}[htbp]
 \begin{minipage}{0.5\hsize}
  \begin{center}
   \includegraphics[width=70mm]{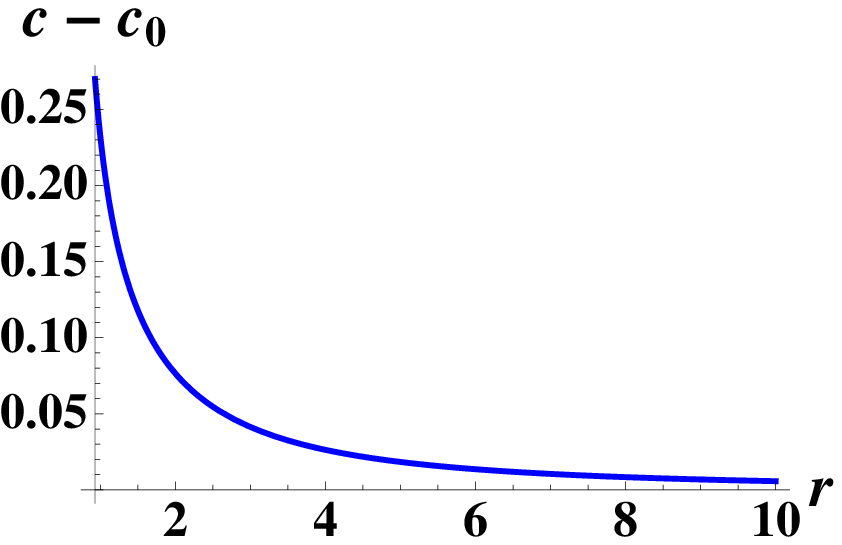}
  \end{center}
  \caption{$c-c_0$ is shown for the parameter choice (I). }
  \label{fig:one}
 \end{minipage}
 \begin{minipage}{0.5\hsize}
  \begin{center}
   \includegraphics[width=70mm]{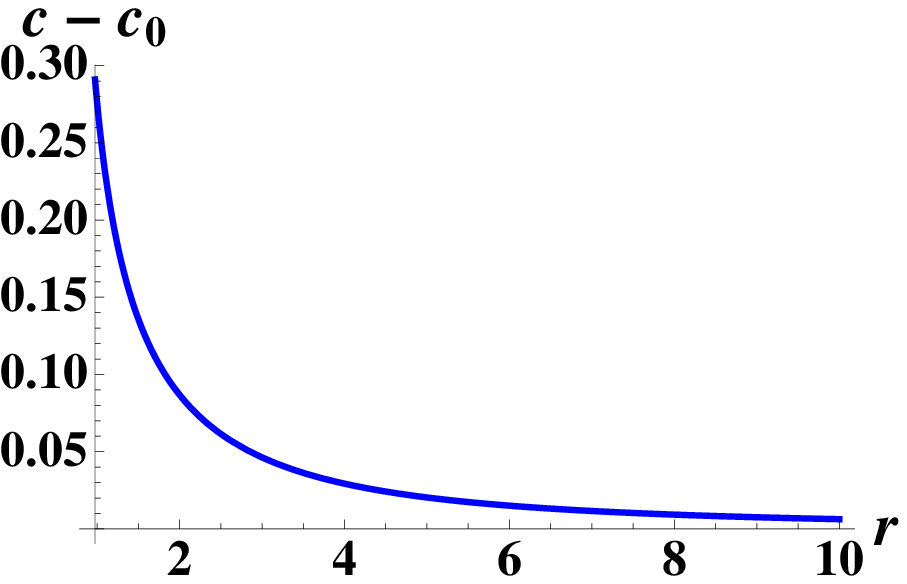}
  \end{center}
  \caption{$c-c_0$ is shown for the parameter choice (II).}
  \label{fig:two}
 \end{minipage}
\end{figure}
\begin{figure}[htbp]
 \begin{minipage}{0.5\hsize}
  \begin{center}
   \includegraphics[width=70mm]{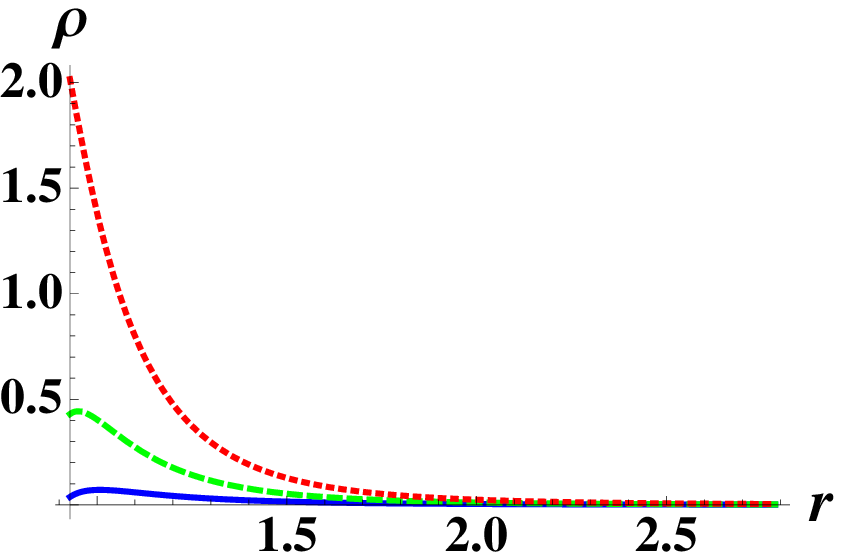}
  \end{center}
  \caption{$\rho=R_{\mu\nu}K^\mu K^\nu~$ for $\gamma=1$~(solid), 
$\gamma=2$~(dashed), and $\gamma=3$~(dotted) are shown for the parameter choice (I). }
  \label{fig:one}
 \end{minipage}
 \begin{minipage}{0.5\hsize}
  \begin{center}
   \includegraphics[width=70mm]{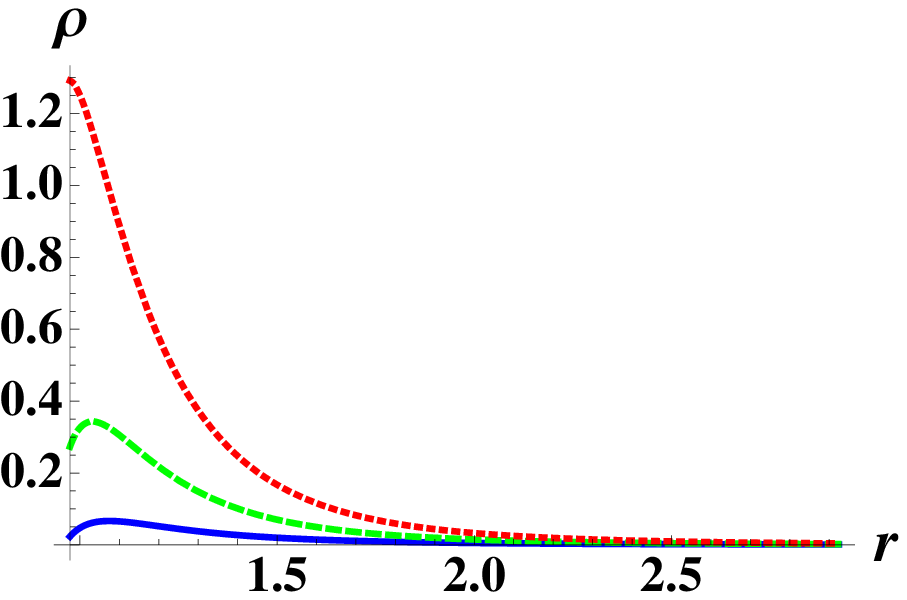}
  \end{center}
  \caption{$\rho=R_{\mu\nu}K^\mu K^\nu~$ for $\gamma=1$~(solid), 
$\gamma=2$~(dashed), and $\gamma=3$~(dotted) are shown for the parameter choice (II).}
  \label{fig:two}
 \end{minipage}
\end{figure}

As shown in Figure 1 - 6, all $v_i$ logarithmically increases as $v_i\sim \ln r$, while $\Omega$ and $c-c_0$ approach zero. 
This implies that the spacetime asymptotically becomes AdS spacetime in the sense mentioned in Sec.~\ref{section3}. 
Since we are interested in the solution satisfying the null energy condition, i.~e., $R_{\mu\nu}K^\mu K^\nu\ge 0$ for any 
null vector $K$, it is worth investigating whether it is satisfied in our solution. Contracting Eq.~(\ref{Einstein_two_vector}) 
with a null vector $K$, we find 
\bea 
\label{null_energy}
& \rho:=R_{\mu\nu}K^\mu K^\nu=\frac{1}{2}(F_{\mu\alpha}{F_\nu}^\alpha+W_{\mu\alpha}{W_\nu}^\alpha) K^\mu K^\nu
\nonumber \\
&+\{a_0+2a_1 (C-C_0)^2\}\{(C-C_0)_\mu K^\mu\}^2. 
\eea  
As the first two terms in the r.~h.~s. are kinetic terms, they would be positive. So, only the term with a negative coefficient $a_1$ 
would be negative. This implies that the null energy condition might be violated when $g(C-C_0,\,K)$ is large. To investigate the sign 
of  Eq.~(\ref{null_energy}), we prepare the following pseudo-orthonormal vectors,  
\begin{align}
\label{tetrad}
& l_+=\p_t-\Omega \p_{x^1}+f\p_r, \nonumber \\
& l_-=\frac{1}{2}\left(\frac{1}{f}\p_t-\frac{\Omega}{f} \p_{x^1}-\p_r\right), \nonumber \\
& E_3=e^{-v_3}\p_{x^1}, \nonumber \\
& E_1=e^{-v_1}(\cos x^1\, \p_{x^2}+\sin x^1\, \p_{x^3}), \nonumber \\
& E_2=e^{-v_2}(\sin x^1\, \p_{x^2}-\cos x^1\, \p_{x^3}) 
\end{align}
satisfying the relation
\begin{align}
& g(l_\pm, l_\pm)=g(l_\pm, E_i)=g(E_i, E_j)=0~(i\neq j), \nonumber \\
& g(l_-,l_+)=-1, \quad g(E_i, E_i)=1. 
\end{align}
Note that $l_+$ and $l_-$ represent outward and inward pointing null vectors, respectively. Thus, 
$l_+$ becomes a tangent vector of the null geodesic generator on the horizon. 

As $E_2$ and $E_3$ are orthogonal to $C-C_0$ vector, we shall consider a null vector $K$  
as
\begin{align}
K=l_++\zeta l_-+\gamma E_1. 
\end{align}
By requiring $g(K,\,K)=0$, we find $\zeta=\gamma^2/2$. In Figure~4, we plot $\rho:=R_{\mu\nu} K^\mu K^\nu$ for several 
parameters $\gamma$ and find that $\rho$ is always positive {for each parameter choice (I) and (II)}. 
This indicates that the null energy condition is satisfied in our numerical solution. 

The Kretschmann scalar curvature invariant $K=R_{\mu\nu\alpha\beta}R^{\mu\nu\alpha\beta}$ can be expanded near the 
horizon $r=r_h$ as a series in $(r-r_h)$. Under the ansatz~(\ref{expansion}), we find that $K$ can be expanded as 
$K=\alpha_0+\alpha_1 (r-r_h)+\cdots<\infty$, where $\alpha_i~(i=0,\,1)$ is a finite coefficient. 
In this sense, the numerical solution is regular near the horizon. As 
mentioned in Sec.~\ref{sec:intro}, this indicates that the black brane solution is not analytic but regular on the horizon 
since the horizon can be compactified, due to the periodicity of the lattice structure. Thus, the solution would be the 
first example of a stationary black hole solution with a compact rotating horizon which evades the rigidity theorem by 
violating one of the crucial conditions, analyticity.

\section{Summary and discussion} 
\label{section6}

In this paper, we have studied in what circumstances a Bianchi black brane 
can have momentum along a direction with translational invariance being broken. 
First, after having given the theorem concerning the decomposition of 
a Killing vector on the horizon, we have shown that in 
the Einstein-Maxwell-dilation theory, 
stationary Bianchi black branes of the type II, VI$_0$, and VII$_0$ 
cannot convey momentum along the direction of no translational invariance.  
For these cases, we have shown under our metric ansatz~(\ref{metric_tr}) 
that the time-asymmetric part of the metric, $\Omega(r)$, must be identically 
zero in the entire region outside the horizon, and thus the geometry 
must be static. 
Two main obstacles for constructing such solutions are the rigidity theorem 
and our choice of boundary conditions: 
The former prevents the horizon from rotating and the latter, 
reflecting the requirement for the absence of non-normalizable mode, 
gives the tight restriction so that $\Omega= 0$ both at the horizon 
and infinity.

Next we have considered Bianchi VII$_0$ black branes in the Einstein-Maxwell 
theory with an additional vector field having a source term. We 
have numerically constructed a solution whose horizon admits 
a translational invariance and thus conveys momentum with non-vanishing 
$\Omega$ on the horizon, which can be viewed as ``rotation'' 
if the horizon cross-sections are compactified.  
Interestingly, the translational invariance is broken just outside 
the horizon. Therefore our solution indicates the existence of a black 
brane solution which is regular but non-analytic at the horizon, 
thereby evading the black hole rigidity theorem. 
Our construction largely relies upon numerical calculation. 
It is hoped to justify by using some analytic methods,  
the existence of such a Bianchi black brane solution with a 
translational invariance being broken in the bulk, but restored 
only on the horizon. 
It is interesting to consider a dual field theory interpretation 
of this solution, and also questions like what is the field theory dual of analyticity of radial direction in bulk. 

\acknowledgments
The work of NI was supported by RIKEN iTHES Project. 
This work was also supported in part by 
JSPS KAKENHI Grant Number 25800143 (NI), 22540299 (AI), 23740200 (KM).

\appendix

\section{Bianchi classification}\label{3dhomogspace}\label{AppA}
The Bianchi classification of 3-dimensional homogeneous spaces is 
in essence a classification of connected, simply connected 3-dimensional 
Lie-groups, which are classified into 9 types by the structure constants 
$C^K{}_{IJ}$ $(I,\;J, \;K=1,\;2,\;3)$ of the corresponding Lie-algebras 
\bea
\left[\xi_I,\xi_J\right] = {C^K}_{IJ} \xi_K \,,   
\eea
with $\{\xi_I \}$ being the generators of the Lie-algebras. Then, one can
choose the invariant basis vectors $\{X_I\}$ so that 
\bea
\left[X_I, X_J\right] = - {C^K}_{IJ} X_K \,, 
\eea 
which can also be expressed in terms of the invariant dual basis one-form 
$\{ \omega^I \}$ as 
\be
\label{diffonef}
d\omega^I = \frac{1}{2} {C^I}_{JK}\, \omega^J \wedge \omega^K \,.
\ee
In this appendix, we provide the structure constants, the generators, the invariant basis,
and the invariant one-forms for the 3 types directly relevant to
our black brane solutions, among 9-types of the Bianchi class. 

\begin{itemize}

\item {\bf Type II}: ${C^1}_{23}=-{C^1}_{32}=1$ and the rest ${C^I}_{JK}=0$ 
\bea
\begin{tabular}{ccc}
$\xi_1=\p_{x^2}$ & $X_1=\p_{x^2}$ & $\omega^1=dx^2-x^1 dx^3$ \\
$\xi_2=\p_{x^3}$ & $X_2=x^1\p_{x^2}+\p_{x^3}$ & $\omega^2=dx^3$  \\ 
$\xi_3=\p_{x^1}+x^3 \p_{x^2}$ & $X_3=\p_{x^1}$ & $\omega^3=dx^1$ \\ 
\label{BianchitypeII}
\end{tabular}
\eea 

\item {\bf Type VI$_0$}: ${C^1}{}_{13}=-C^1{}_{31}=1$,
      ${C^2}_{23}=-{C^2}_{32}=-1$
      and the rest ${C^I}_{JK}=0$
 \bea
 \begin{tabular}{cccc}
 $\xi_1=\p_{x^2}$ & $X_1=e^{x^1}\p_{x^2}$ & $\omega^1=e^{-x^1}dx^2$  \\
 $\xi_2=\p_{x^3}$ & $X_2=e^{-x^1}\p_{x^3}$ & $\omega^2=e^{x^1}dx^3$
          \\ 
 $\xi_3=\p_{x^1}+x^2 \p_{x^2}-x^3 \p_{x^3}$ & $X_3=\p_{x^1}$ &
          $\omega^3=dx^1$
            \\ 
 \label{BianchitypeVI}
 \end{tabular}
 \eea

\item {\bf Type $\bf{VII_0}$}: ${C^1}_{23}=-{C^1}_{32}=-1$, ${C^2}_{13}=-{C^2}_{31}=1$ and the 
 rest ${C^I}{}_{JK}=0$.
 \bea
 \begin{tabular}{cc}
 $\xi_1 = \p_{x^2}$ \! & $X_1=\cos(x^1)\p_{x^2}+\sin(x^1)\p_{x^3}$ \\
 $\xi_2 = \p_{x^3}$ \!& $X_2=-\sin(x^1)\p_{x^2}+\cos(x^1)\p_{x^3}$ \\
 $\xi_3=\p_{x^1}-x^3 \p_{x^2}+x^2 \p_{x^3}$ \!&  $X_3=\p_{x^1}$ \\
 \label{BianchitypeVII0}
 \end{tabular}
\eea
and 
\bea
&& \omega^1=\cos(x^1)dx^2+\sin(x^1)dx^3 , \nonumber \\ 
&&\quad
  \omega^2=-\sin(x^1)dx^2+\cos(x^1)dx^3 , \quad \omega^3=dx^1   \,.
\label{BianchitypeVII0}
\eea

\end{itemize}



\end{document}